\documentstyle[psfig,aaspp4]{article} 

\newcommand{\msun}{\mbox{${\rm M}_\odot$}}

\newcommand{\nbody}{\mbox{{{\em N}-body}}}

\newcommand{\trlx}{\mbox{${t_{\rm rt}}$}}
\newcommand{\trlxo}{\mbox{${t_{\rm rt,0}}$}}
\newcommand{\trxh}{\mbox{${t_{\rm rh}}$}}

\newcommand{\tdf}{{\mbox{${t_{\rm df}}$}}}
\newcommand{\tdiss}{\mbox{${t_{\rm diss}}$}}

\newcommand{\mto}{\mbox{${m_{\rm to}}$}}

\newcommand{\mmin}{\mbox{$m_{\rm min}$}}
\newcommand{\mm}{\mbox{$\langle m \rangle$}}

\newcommand{\rhm}{\mbox{${r_{\rm hm}}$}}
\newcommand{\Rgc}{\mbox{${R}$}}

\newcommand{\rJ}{\mbox{${r_{\rm J}}$}}
\newcommand{\mJ}{\mbox{${m_{\rm J}}$}}
\newcommand{\nJ}{\mbox{${n_{\rm J}}$}}

\newcommand{\vorb}{\mbox{$v_{\rm c}$}}

\newcommand{\erf}{\mbox{${\rm erf}$}}

\newcommand{\xgc}{\alpha}
\newcommand{\ximf}{x}
\def\apgt{\ {\raise-.5ex\hbox{$\buildrel>\over\sim$}}\ }
\def\aplt{\ {\raise-.5ex\hbox{$\buildrel<\over\sim$}}\ }

\begin{document}


\title{The fate of star clusters near the Galactic center I: 
	Analytic considerations}

\author{Stephen L.\ W.\ McMillan$^{1}$,
	Simon F.\ Portegies Zwart$^{2, 3, \star}$
}

\noindent
$^{1}$ Department of Physics,
		 Drexel University,
                 Philadelphia, PA 19104, USA \\
$^{2}$ Astronomical Institute `Anton Pannekoek', 
		University of Amsterdam, Kruislaan 403 \\
$^{2}$ Institute for Computer science,
		 University of Amsterdam, Kruislaan 403 \\

$^\star$ SPZ is a KNAW Fellow \\

\lefthead{Portegies Zwart et al.}
\righthead{Life and death of star clusters}

\begin{abstract}

A star cluster in a galactic nucleus sinks toward the galactic center
due to dynamical friction.  As it spirals inward, the cluster loses
mass due to stellar evolution, relaxation driven evaporation, and
tidal stripping, eventually dissolving in the galactic tidal field.
We model the inspiral of dense young star clusters near the center of
our Galaxy to study the extent of the region of parameter space in
which the cluster can reach the inner parsec of the Galaxy within a
few million years.  Since we neglect changes in cluster structure due
to internal evolution, the present study is most applicable to star
clusters less than about one initial relaxation time old.  We find
that only star clusters with initial masses $\apgt 10^5$\,{\msun} can
reach the Galactic center from an initial distance of $\apgt 60$\,pc
within one initial relaxation time or a few million years, whichever
is smaller.

\keywords{stellar dynamics ---
	  methods: analytical ---
	  Galaxy: Bulge ---
	  Galaxy: Nucleus ---
	  Galaxy: Stellar content ---
	  open clusters and associations: individual IRS 16 ---
}
\end{abstract}

\section{Introduction}
The innermost $\sim100$\,pc of the Milky Way Galaxy contains a number
of intriguing objects.  These include the central $\sim 2-3\times
10^6$\,{\msun} black hole (Genzel et al.\, 2000; Ghez et
al.\,2000),\nocite{2000MNRAS.317..348G}\nocite{2000Natur.407..349G} a
cluster containing at least 15 massive young stars (Tabmley \& Rieke
1993; Krabbe et
al.\,1995),\nocite{1993ApJ...414..573T}\nocite{1995ApJ...447L..95K} a
much larger population of older stars (Alexander
1999),\nocite{1999ApJ...527..835A} and at least two young dense star
clusters---the Arches and the Quintuplet systems (Nagata et al.\,1995;
Nagata et al.\ 1990; Okuda et al.\, 1990).
\nocite{1995AJ....109.1676N}\nocite{1990ApJ...351...83N}\nocite{1990ApJ...351...89O}


Krabbe et al.\,(1995) found $\sim15$ bright He I emission line stars
in the Galactic center.  They are part of the co-moving 7--8\,Myr old
$\apgt 10^4$\,{\msun} association known as IRS\,16 (Tamblyn \& Rieke
1993; Krabbe 1995)\nocite{1993ApJ...414..573T}, and are accompanied by
many less luminous stars of spectral types O and B (Genzel et
al.\,2000). Detailed spectroscopic analysis of the Galactic center
region (Najarro et al.\,1994) indicate that these emission-line stars
are evolved, with a high surface ratio of helium to hydrogen $n_{\rm
He}/n_{\rm H} = 1$ to 1.67.  Allen et
al.\,(1990)\nocite{1990MNRAS.244..706A} classify them as Ofpe/WN9
stars, while Najarro et al.\,(1997)\nocite{1997A&A...325..700N}
identify them as 60--100\,{\msun} Luminous Blue Variables (LBVs), the
late evolutionary stages of very massive stars (Langer et
al.\,1994).\nocite{1994A&A...290..819L} Depending on the
interpretation of the data, the age of IRS\,16 therefore lies in the
range 3--7\,Myr, the lower figure corresponding to the LBV
identification.

One possible explanation for these stars is a recent $\sim
10^4\,\msun$ starburst (Krabbe et al.\,1995).  However, this model is
problematic, as the formation of stars within a parsec of the Galactic
center is difficult; the Galactic tidal field is sufficient to unbind
gas clouds with densities $\aplt 10^7$\,cm$^{-3}$ (G\"usten \& Downes
1980).  Gerhard (2001) has proposed that a million-solar-mass star
cluster formed at a distance of $\aplt 30$\,pc from the Galactic
center could have reached the Galactic center via dynamical friction
before being disrupted by the Galactic tidal field or by internal
dynamical evolution.  This qualitative argument solves the problem of
the presence of young, very massive stars in the Galactic center.
Gerhard's dynamical friction time scale assumed that the stellar
density in the vicinity of the Galactic center is described by an
isothermal sphere; in addition, he ignored stellar mass loss and the
internal dynamical evolution of the cluster.  In this paper we present
a more quantitative approach to the problem.

This is the first in a series of papers in which we consider the time
scale on which a star cluster sinks to the Galactic center and is
disrupted by the Galactic tidal field.  In the semi-analytic
calculations presented here, we study the inspiral of three quite
different cluster models.  We begin with the simplifying approximation
that the inspiraling object has constant mass.  Later, we relax that
assumption and allow the cluster to lose mass, first by tidal
stripping, then also by stellar evolution and relaxation.  For
definiteness, and for purposes of illustration, we adopt a simple
analytic prescription for mass loss from the evolving cluster, and
investigate its consequences.  In a future paper we will incorporate
more realistic treatments of cluster dynamics.

The organization of this paper is as follows. In \S2 we first consider
the orbital decay of a unevolving point mass.  In \S3 we expand our
study to include clusters of nonzero radii, allowing their masses to
evolve in time as material is stripped by the Galactic tidal field.
The introduction of physical parameters into our models then allows us
to incorporate simple treatments stellar mass loss and evaporation
within our simple model. In \S4 we apply the model to star clusters
near the Galactic center, to determine the region of parameter space
in which clusters can transport a considerable fraction of their
initial mass to within a few parsecs of the Galactic center before
disruption.  We discuss our results and conclude in \S5.


\section{Inspiral with constant mass}\label{sect:constantmass}
We begin our study with the simplifying assumption that the mass of
the inspiraling object is constant.  This idealization may be
appropriate for a single massive black hole or a very compact star
cluster which is much smaller than its Jacobi radius, the limiting
radius of a cluster in the tidal field of the Galaxy.  In the latter
case, however, for the constant-mass approximation to hold, internal
dynamical evolution of the cluster should also be negligible on the
time scale on which the cluster sinks to the Galactic center. In
practice, especially for small clusters, this will not be the case, as
we discuss in \S\ref{sect:discussion}.

\subsection{Dynamical friction}\label{sect:dynfric}
We characterize the mass $M$ within a sphere with radius $\Rgc$ centered
on the Galactic center as a power law:
\begin{equation}
	M(\Rgc) = A\Rgc^\xgc\,,
\label{Eq:Galaxymass}\end{equation}
where $A$ and $\xgc$ are constants, with $1<\xgc<2$ of interest here.
The density at distance $\Rgc$ then is
\begin{equation}
	\rho(\Rgc) 
	        = {A\xgc \over 4\pi} \Rgc^{\xgc-3}\,,
\label{Eq:rho}\end{equation}
and we can write down expressions for the orbital acceleration at
distance
$\Rgc$ from the Galactic center
\begin{equation}
	a(\Rgc) 
		= GA\Rgc^{\xgc-2},
\label{Eq:acceleration}\end{equation}
the potential
\begin{equation}
	\phi(\Rgc) 
		= {GA \over \xgc-1} \Rgc^{\xgc-1},
\end{equation}
the circular velocity
\begin{equation}
	\vorb^2(\Rgc) 
		= GA\Rgc^{\xgc-1},
\label{Eq:vorbit}\end{equation}
and the total energy of a circular orbit
\begin{equation}
	E_c(\Rgc) 
	       = {\textstyle\frac12} GA\Rgc^{\xgc-1}
	\left( {\xgc+1 \over \xgc-1} \right).
\label{Eq:Energy}\end{equation}

The object's acceleration due to dynamical friction is (Binney \&
Tremaine 1987, p.\,425)
\begin{equation}
	{\bf a}_f = - 4\pi\ln\Lambda\,G^2\rho\,m
			{{\bf v}_c \over \vorb^3}\,\chi\,.
\label{Eq:afric}\end{equation}
Here, $m$ is the mass of the object, ${\bf v}_c$ is its velocity
vector (in a circular orbit around the Galactic center), $\ln\Lambda
\sim \ln\langle r\rangle/\Rgc \sim 5$ is the Coulomb logarithm (where
$\langle r\rangle$ is the object's characteristic radius---roughly the
half-mass radius in the case of a cluster), $G$ is the gravitational
constant, and
\begin{equation}
	\chi \equiv \erf(X) - {2X \over \sqrt{\pi}}e^{-X^2},
\label{Eq:chi}\end{equation}
where $X = \vorb/\sqrt{2}\sigma$ and $\sigma^2(\Rgc)$ is the local
one-dimensional velocity dispersion, assumed isotropic.

Substitution of Eqs.\,(\ref{Eq:Galaxymass}), (\ref{Eq:rho}) and
(\ref{Eq:vorbit}) into Eq.\,(\ref{Eq:afric}) results in
\begin{equation}
	a_f \equiv |{\bf a}_f| = \xgc \chi \ln\Lambda \frac{G m }{\Rgc^2}\,,
\label{Eq:Fdf_II}\end{equation}
from which we note that
\begin{equation}
	{a_f \over a(\Rgc)} = \xgc \chi \ln\Lambda {m \over M}.
\end{equation}
For $\xgc=1.2$, we obtain $X \simeq 0.89$ (see Appendix A), and hence
$\chi \simeq 0.34$.  With $\ln\Lambda = 5$ we find $a_f/a(\Rgc) \simeq
2m/M$.

\subsection{Orbital decay}\label{sect:timescale}
We can now derive the inspiral time scale for a star cluster with
constant mass $m$ in a power-law density profile given by
Eq.\,(\ref{Eq:Galaxymass}).  The time derivative of
Eq.\,(\ref{Eq:Energy}) is
\begin{eqnarray}
	{dE_c \over dt}
		&=& {\textstyle\frac12}
				 (\xgc+1) G A \Rgc^{\xgc-2} {d\Rgc \over dt}
							 	\nonumber \\
		&=& -\chi \ln\Lambda \, G^2 {\rho m \over \vorb}\,,
\end{eqnarray}
where the second equation expresses the work done by dynamical
friction (Eq. \ref{Eq:Fdf_II}).  Hereafter, $\Rgc$ should be
interpreted as $\Rgc(t)$, the distance from the cluster in question to
the Galactic center.  Substitution of Eqs.\,(\ref{Eq:rho}) and
(\ref{Eq:vorbit}) leads to
\begin{equation}
	{dE_c \over dt} = - \xgc \chi \ln\Lambda G^{3/2} A^{1/2} m 
				 \Rgc^{(\xgc-5)/2},
\end{equation}
whence
\begin{equation}
	{d\Rgc \over dt} = - \gamma \Rgc^{-(\xgc+1)/2}\,,
\label{Eq:dRdt}\end{equation}
where
\begin{equation}
	\gamma = 2m\ln\Lambda {\xgc \chi \over \xgc+1}
			\left({G \over A}\right)^{1/2}.
\label{Eq:gamma}\end{equation}

Solving Eq.\,(\ref{Eq:dRdt}) with $R(t) = R_0$ at time $t = 0$ results
in
\begin{equation}
	R(t) = R_0 \left[
			1 - {(\xgc+3)\gamma \over 2R_0^{(\xgc+3)/2}}\,
			t
		   \right]^{2/(\xgc+3)}.
\end{equation}
Setting $R = 0$ at $t=\tdf$ and substituting
Eq.\,(\ref{Eq:Galaxymass}) yields
\begin{eqnarray}
	\tdf	
	     &=& {\xgc+1 \over \xgc(\xgc+3)} {1 \over \chi \ln\Lambda}
		 \left( {M_0 \over G} \right)^{1/2}
		 {R_0^{3/2} \over m}\,,
\label{Eq:tdf}\end{eqnarray}
where $M_0 = M(R_0)$.  In terms of the orbital period of a circular
orbit around the Galactic center at distance $R_0$, $T_0 =
2\pi\left(GM_0/R_0^3\right)^{-1/2}$, Eq.\,(\ref{Eq:tdf}) becomes
\begin{equation}
	{\tdf \over T_0} = {\xgc+1 \over 2 \pi \xgc(\xgc+3)} 
		{1 \over \chi \ln\Lambda} 
		{M_0 \over m}\,.
\label{Eq:tdftorb}\end{equation}
For $\xgc =1.2$, $M_0/m = 10^3$, and $\ln\Lambda = 5$, we find $\tdf
\simeq 40\,T_0$.


\section{Clusters with variable mass}\label{sect:variablemass}
We now consider the possibility that the mass of the cluster varies
with time: $m = m(t)$.  Most mass loss from the cluster is the result
of tidal stripping as the cluster sinks toward the Galactic center.
We begin by determining the Jacobi (tidal) radius $\rJ$ of the cluster
in the tidal field of the Galaxy.

\subsection{Mass of a tidally limited cluster}\label{Sect:clustermass}
The differential acceleration at distance $\rJ$ from the center of the
cluster is obtained from Eq.\,(\ref{Eq:acceleration}):
\begin{equation}
	\Delta a_{\rm tide} \approx 
		(\xgc-2) GA R^{\xgc-3}\rJ\,,
\label{Eq:dudR}\end{equation}
or, relative to the internal cluster acceleration at $\rJ$,
\begin{equation}
	{|\Delta a_{\rm tide}| \over a_{\rm J}} 
		= (\xgc-2) \left({M \over \mJ}\right)
			\left({\rJ \over R}\right)^3\,.
\end{equation}
Here $a_{\rm J} = G\mJ/\rJ^2$ and $\mJ$, the cluster mass within
radius $\rJ$ (still to be determined), will henceforth be identified
as the cluster mass.  Setting $|\Delta a_{\rm tide}| = a_{\rm J}$, we
find
\begin{equation}
	\left({M \over \mJ}\right) 
	\left({\rJ \over R}\right)^3 = {1 \over 2-\xgc}\,.
\end{equation}
This may be conveniently (and conventionally) expressed in terms of
average densities $\overline{\rho}_{\rm J} = 3\mJ/4\pi\rJ^3$ and
$\overline{\rho}_{\rm G} = 3M/4\pi R^3$, as
\begin{equation}
	\overline{\rho}_{\rm J} = (2-\xgc)\, \overline{\rho}_{\rm G}\,.
\label{Eq:density}\end{equation}

To proceed further, we must make a connection between $\mJ$ and $\rJ$.
Two particularly simple cluster density profiles lend themselves
easily to analytic development:
\begin{enumerate}

\item A homogeneous sphere of mass $m_0$, radius $b$, and uniform
density
\begin{equation}
	\rho_0 = {3m_0 \over 4\pi b^3}\,.
\label{Eq:rho0}\end{equation}

\item A Plummer (1911)\nocite{1911MNRAS..71..460P} model of mass $m_0$
and scale radius $b$, with
\begin{equation}	
	\rho(r) = \rho_0 \left( 1 + r^2/b^2 \right)^{-5/2},
\label{Eq:Plummer}\end{equation}
where $\rho_0$ is again given by Eq.\,(\ref{Eq:rho0}).

\end{enumerate}
Note that, in each case we assume {\em fixed} parameters $m_0$ and
$b$---that is, we neglect structural changes in the cluster due to
dynamical evolution or stellar mass loss.  This assumption greatly
simplifies the calculation, but clearly is of questionable validity
when the internal dynamical time scales are comparable to the inspiral
time (see \S\ref{sect:discussion}).  In the next subsection we expand
our model to allow for the effects of mass loss due to stellar
evolution and escaping stars.  A more complete treatment of the
cluster's structural evolution will be the subject of a future paper.

For the homogeneous sphere (case 1), the desired relation between
$\mJ$ and $\rJ$ is simple:
\begin{equation}
	\mJ = \left\{
		\begin{array}{ll}
			m_0 \left(\rJ/b \right)^3 & ~~~~~(\rJ<b),\\
			m_0 & ~~~~~(\rJ\ge b).
		\end{array}
	      \right.
\end{equation}
No solution to Eq.\,(\ref{Eq:density}) exists for
$\overline{\rho}_{\rm G} > \rho_0/(2-\xgc)$, and the cluster is
destroyed at Galactocentric radius
\begin{equation}
	R_{\rm min} = \left[{m_0 \over
				(2-\xgc)Ab^3}\right]^{1/(\xgc-3)}.
\end{equation}
Outside $R_{\rm min}$, $\rJ > b$ and $m=m_0$.  Inside, $\rJ = m = \mJ
= 0$.

For the Plummer model (case 2),
\begin{equation}
	\overline{\rho}_{\rm J} = \rho_0 
				  \left(1 + \rJ^2/b^2\right)^{-3/2}
\end{equation}
and  again, no solution exists for $R<R_{\rm min}$.  Outside $R_{\rm
min}$,
$\rJ(R)$ satisfies
\begin{eqnarray}
	1 + \rJ^2/b^2 
		      &=& (2-\xgc)^{2/3}\,\left( {m_0 \over M_0} \right)^{2/3} 
			      \left( {R_0 \over b} \right)^2
			      \left( {R \over R_0} \right)^{2 - 2\xgc/3}\,.
\end{eqnarray}
The mass of the cluster is then given by Eq.\,(\ref{Eq:density}):
\begin{eqnarray}
	\mJ(R) 
	       &=& (2-\xgc) \, M_0 \left( {\rJ \over R_0} \right)^3
		    	 \left( {R \over R_0} \right)^{\xgc-3}.
\label{Eq:mJ}\end{eqnarray}
We use this model as the basis for our discussion in the remainder of
the paper.

\subsection{Mass loss from stellar evolution}
Many clusters dissolve so quickly that stellar evolution barely
affects their mass.  However, if the cluster survives for more than a
few million years, mass loss from the most massive stars may become
important (see McMillan 2003 for a recent review).  Recent detailed
{\nbody} simulations by Portegies Zwart et al.\,(2001) have quantified
the expansion of a tidally-limited cluster as its mass decreases.  The
expansion drives more rapid disruption, while the mass loss slows the
inspiral.

We include stellar mass loss in our model as follows.  First we
rewrite Eq.\,(\ref{Eq:mJ}) as $\mJ(R) = \mm \nJ(R)$, where {\nJ(R)} is
the number of stars within the Jacobi radius and {\mm} is the mean
stellar mass, which is now a function of time due to stellar
evolution.  We assume that the mass functions of the cluster and of
the escaping stars are identical.  (Again, this is equivalent to the
neglect of internal dynamical evolution.)  We parametrize the
cluster's expansion in response to stellar mass loss by
\begin{equation}
	b = b_0\,\mm_0/\mm,
\end{equation}
which is equivalent to the assumption that the cluster loses mass
adiabatically, as found by Portegies Zwart et al.\,(2001).

The mean mass in the cluster can be computed from the initial mass
function.  For clarity we assume that all the mass in stars having
masses above the cluster's turn-off mass is simply lost from the
cluster.  So long as the turn-off mass exceeds $\sim 8$\,\msun\, this
assumption is justified by the high-velocity kick imparted to compact
objects by the supernovae in which they form, allowing them to escape
from the cluster.  For older clusters this assumption breaks down as
lower-mass stars turn into white dwarfs, which do not receive high
velocities at formation, although such clusters are not of direct
interest in the present paper.  Integrating the initial mass function,
we find
\begin{equation}
	\mm =  \left({1-\ximf \over 2-\ximf}\right)\,
	    	    {\mto^{2-\ximf} - \mmin^{2-\ximf}
			\over 
		     \mto^{1-\ximf} - \mmin^{1-\ximf}}\,.
\label{Eq:mm}\end{equation}
Here $\ximf$ is the exponent for the (assumed) power-law mass function
(Salpeter: $\ximf = 2.35$), {\mto} is the cluster turn-off mass, and
{\mmin} is the lower mass limit.  We determine the turn-off mass using
fits to the stellar evolution models of Eggleton, Fitchet \& Tout
(1998).\nocite{1989ApJ...347..998E}

\subsection{Mass loss due to relaxation}
A tidally limited star cluster in the tidal field of the Galaxy will
also lose mass due to internal relaxation as occasional interactions
between cluster members result in velocities high enough for stars to
escape the cluster potential.  Portegies Zwart \& McMillan
(2002)\nocite{2002ApJ...565..265P} have simulated star clusters near
the Galactic center; they derive the following approximate expression
for relaxation-driven cluster mass loss:
\begin{equation}
	m(t) = m_0 \left(1 - {t \over 0.29 \trlx} \right).
\label{Eq:relaxation}\end{equation}
Here {\trlx} is the relaxation time at the cluster tidal
radius,
\begin{equation}
	\trlx = 2.05\,{\rm Myr}
		\left( {\rJ \over 1\,{\rm pc}}	\right)^{3/2} 
		\left( {\mJ \over \msun}	\right)^{-1/2} 
		{\nJ \over \log(0.4 \nJ)}\,,
\label{Eq:trlx}\end{equation}
where $n_J$ is the number of stars contained within the Jacobi radius.

The clusters in the study of Portegies Zwart \& McMillan (2002) did
not spiral in to the Galactic center, so the relaxation time at the
tidal radius remained constant over the lifetime of the cluster.  In
our case, where clusters sink toward the Galactic center, the
relaxation time at the tidal radius changes with time.  We therefore
recast Eq.\,(\ref{Eq:relaxation}), as follows.  Differentiating
Eq.\,(\ref{Eq:relaxation}) with respect to time, identifying $m(t)$
with $\mJ$ and $\trlx$ with the instantaneous relaxation time at
$\rJ$, and including the radial dependence of the relaxation time,
assuming a tidally limited cluster, we obtain
\begin{equation}
	{dm \over dt} = -{m_0 \over 0.29\trlxo} 
			\left( {R \over R_0} \right)^{(\xgc-3)/2}.
\label{Eq:relaxation_diff}\end{equation}

For the purposes of this paper, we draw a distinction between the
processes of {\em tidal stripping}, in which stars outside the Jacobi
radius are removed by the Galactic tidal field as the cluster sinks
toward the Galactic center and the Jacobi radius shrinks, and {\em
evaporation}, in which stars are driven across the instantaneous
Jacobi radius by internal two-body relaxation.  All models discussed
in the following section include tidal stripping; models discussed in
\S\ref{sect:SE} and subsequently also include both evaporation-driven
and stellar-evolution mass loss.


\section{Results}
From \S\ref{sect:timescale}, the distance from the cluster to the
Galactic center satisfies
\begin{equation}
	{dR \over dt} = -\gamma(R)\, R^{-(\xgc+1)/2}
\end{equation}
(Eqs.\,\ref{Eq:dRdt} and \ref{Eq:gamma}), with $R=R_0$ at $t=0$.
Transforming to dimensionless variables $\xi = R/R_0$ and $\tau =
t/T_0$, and substituting Eq.\,(\ref{Eq:gamma}), we rewrite this
equation in the form
\begin{equation}
	{d\xi \over d\tau} = {4\pi \xgc \over \xgc+1}
				\chi \ln\Lambda
				{\mJ \over M_0} \xi^{-(\xgc+1)/2}.
\label{Eq:dxidtau}\end{equation}
For a Plummer model, the cluster mass {\mJ} varies as a function of
$R$ and therefore $\xi$ via Eq.\,(\ref{Eq:mJ}).  We solve
Eq.\,(\ref{Eq:dxidtau}) numerically, as it admits no simple analytic
solution.  For all models we adopt $A =4.25\times10^6$\,{\msun} and
$\xgc=1.2$ (Sanders \& Lowinger 1972, Mezger et
al.\,1999).\nocite{1972AJ.....77..292S}\nocite{1999A&A...348..457M}

For simplicity we assume that $\chi\ln\Lambda = 1$ for the remainder
of this section, unless indicated otherwise.  A value of
$\chi\log\Lambda = 1.2$ or 1.3 is probably more appropriate (Binney \&
Tremaine 1987; Spinnato, Fellhauer \& Portegies Zwart 2003).  The
dynamical friction time scale is inversely proportional to
$\chi\ln\Lambda$ (see Eq.\,\ref{Eq:tdf}), so the effects of different
choices can be easily estimated.

\subsection{Solutions without stellar evolution}
\begin{figure}[htbp!]
\psfig{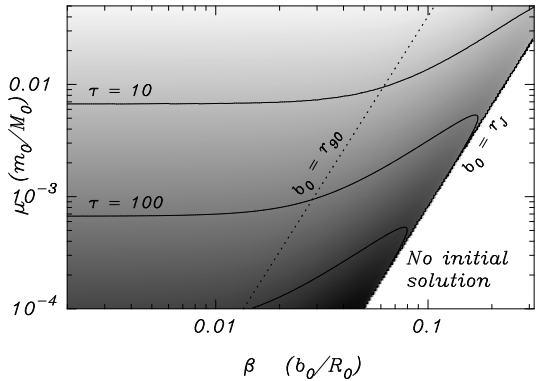}
\caption[]{Scaled cluster lifetime (contours and gray shades) as a
function of the dimensionless parameters $\beta$ and $\mu$ (see
Eqs.\,\ref{Eq:dxidtau} and \ref{Eq:au}), for models with tidal
stripping but without additional mass loss by stellar evolution or
evaporation.  The parameter $\beta$ is the ratio of the initial
cluster length scale to the initial distance to the Galactic center.
The parameter $\mu$ is the ratio of the initial cluster mass to the
mass of the Galaxy contained within the initial orbit.  The numerical
labels on the contours give the disruption time $\tau$ in units of the
cluster's initial orbital period around the Galactic center.  The gray
shades provide the same information as the contours; darker shades
represent longer cluster lifetime.  The dotted line indicates the
values of $\beta$ and $\mu$ corresponding to $b_0=0.9\rJ$: the
characteristic scale of the initial Plummer model is 90\% of the
cluster Jacobi radius.  No initial solution exists for the area to the
right of the curve $b_0=\rJ$ (also indicated). }
\label{Fig:au_cnt}
\end{figure}

For systems without significant stellar mass loss or evaporation, the
evolution may be conveniently parametrized by the dimensionless
quantities
\begin{eqnarray}
	\beta &=& b_0/R_0, \nonumber\\
	\mu &=& m_0/M_0.
\label{Eq:au}\end{eqnarray}
The contours and gray shades in Fig.\,\ref{Fig:au_cnt} present the
dissolution time ($\tau$) of the cluster as a function of $\beta$ and
$\mu$.

Fig.\,\ref{Fig:au_cnt} shows that compact, massive clusters have the
shortest lifetimes, that the lifetime decreases with increasing mass
at fixed initial cluster radius ($b_0$), increases with increasing
radius at fixed mass, and is largely independent of the radius for
small radii.  This last point simply means that clusters initially
well inside their Jacobi radii ($b_0\ll0.9\rJ$) experience significant
stripping only near the end of the inspiral process.  There is no
initial solution when $\overline{\rho}_{\rm G} > \rho_0/(2-\xgc)$,
i.e.~when $b_0 > \rJ$.

\begin{figure}[htbp!]
\psfig{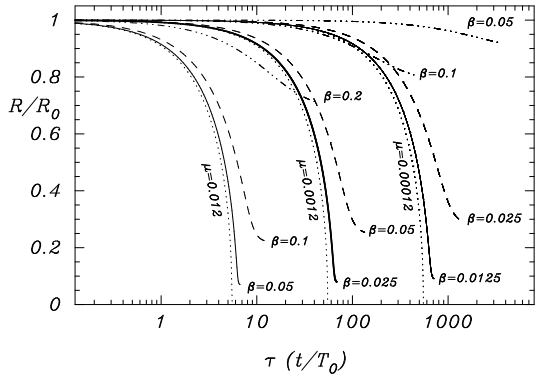}
\caption[]{Scaled distance to the Galactic center $R/R_0$ as a
function of scaled time $t/T_0$, for selected combinations of $\beta$
and $\mu$.  The three families of models shown have $\mu=0.012$ (left
set of curves), $\mu=0.0012$ (middle), and $\mu=0.00012$ (right-most
curves).  The dotted lines give the evolution for a constant point
mass ($\beta=0$); other curves present models with $\beta$ as
indicated.  The model corresponding to $\mu=0.0012, \beta=0.025$
(heavy solid line) is the basis for
Figs.\,\ref{Fig:RtNXk}--\ref{Fig:5}.}
\label{Fig:F2}
\end{figure}

Since the stellar density diverges toward the Galactic center, no
extended cluster can actually reach $R=0$ (although a black hole can).
Fig.\,\ref{Fig:F2} shows the cluster's distance to the Galactic center
as a function of time (again in units of the initial orbital period of
the cluster around the Galactic center) for several selected values of
$\beta$ and $\mu$.  Not surprisingly, more massive clusters (larger
values of $\mu$) spiral in more quickly, and physically larger
clusters (larger $\beta$) dissolve at larger distances from the
Galactic center.

The long lifetimes of clusters with $b_0 \apgt 0.9\rJ$ (see
Fig.\,\ref{Fig:au_cnt}) and small values of $\mu$ are due to the weak
effect of dynamical friction in those cases.  Since we ignore stellar
mass loss and internal dynamical evolution (specifically, evaporation)
in this simple model, such clusters survive for unrealistically long
times.  In practice, these systems will be strongly affected by
stellar evolution and evaporation, as we now demonstrate.

\subsection{Evolution with stellar mass loss and relaxation}\label{sect:SE}

By selecting the Galactic center as representative nucleus we can
attach physical units to the selected values of $\mu$ and $\beta$.
The advantage of introducing physical parameters is that the numbers
become more intuitive, but of course we lose the scale-free solution
from previous section.  Another advantage of fixing the scaling is
that we can take stellar evolution and internal relaxation into
account.  Stellar mass loss (via Eq.\,\ref{Eq:mm}) and evaporation
(via Eq.\,\ref{Eq:relaxation_diff}) are included by solving
Eq.\,(\ref{Eq:dxidtau}).  For most calculations we adopted a Salpeter
initial mass function between 0.1\,{\msun} and 100\,\msun.  The effect
of relaxing this assumption is illustrated in Fig.\,\ref{Fig:5} below.

Fig.\,\ref{Fig:RtNXk} shows distances to the Galactic center as
functions of time of model clusters having initial masses of
64,000\,{\msun} (a) and 256,000\,{\msun} (b).  For each selected
initial distance ($R_0=2$\,pc, 4\,pc, 8\,pc and 16\,pc) we choose a
range of initial values for the cluster scale $b_0 = 0.2$\,pc, 0.4,
and 0.8\,pc.  The choice of $m_0=64,000\,\msun$, $R_0=8$\,pc, and
$b_0=0.2$\,pc corresponds to the ``standard'' model indicated in
Fig.\,\ref{Fig:F2}.  These models were computed taking both stellar
mass loss and evaporation into account.  Models with $b_0=0$ (point
mass case, without stellar mass loss or evaporation) are also included
for comparison; they are identical to the calculations presented in
Fig.\,\ref{Fig:F2}.

\begin{figure}[htbp!]
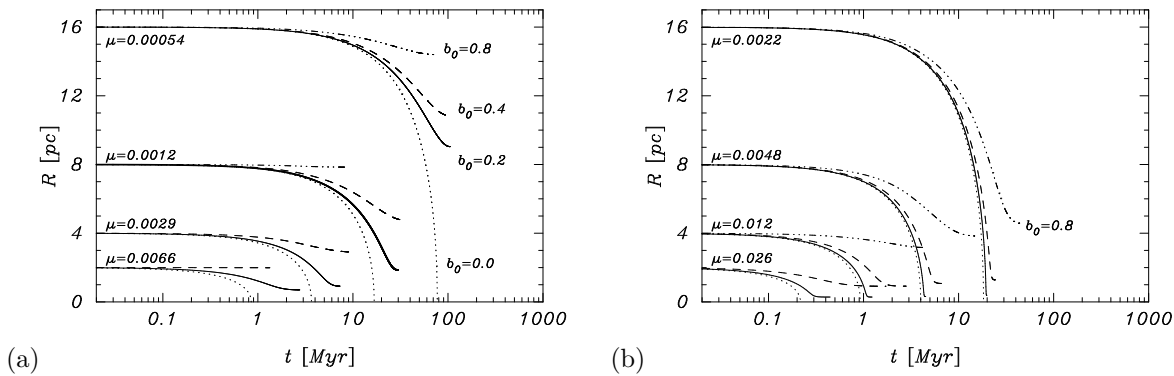

(a)\psfig{figure=f3a.ps,width=7.cm,angle=-90}~~~~
(b)\psfig{figure=f3b.ps,width=7.cm,angle=-90}
\caption[]{Time-evolution of Galactocentric distance $R$ for models
with (a) $m=64,000$\,{\msun} and (b) 256,000\,\msun, assuming
$\chi\ln\Lambda = 1$.  Initial cluster scales $b_0$ are 0.0 (dotted
lines), $b=0.2$ (solid line), 0.4 (dashes) and 0.8 (dash-3-dotted
lines).  The corresponding values for $\mu$ are indicated near the
start of each family of curves.  The ``standard'' model marked in
Fig.\,\ref{Fig:F2} is also indicated here (heavy solid line).  For
reference, the orbital periods of clusters at 2\,pc, 4\,pc, 8\,pc and
16\,pc from the Galactic center are 0.09\,Myr, 0.16\,Myr, 0.30\,Myr
and 0.56\,Myr, respectively.  }
\label{Fig:RtNXk}
\end{figure}

\begin{figure}[htbp!]
\psfig{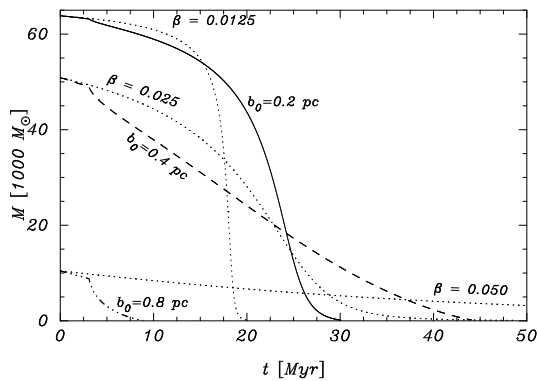}
\caption[]{Mass as a function of time for a 64,000\,{\msun}
star cluster at a distance of 8\,pc
from the Galactic center ($\mu = 0.00012$)  with
various values for $b_0$ (as indicated).  The dotted curves present
the corresponding evolution of the cluster without stellar evolution
or evaporation.  The values of $\beta$ correspond to the choices of
$b_0$.  All calculations were made using $\chi \ln\Lambda = 1$.  }
\label{Fig:6}
\end{figure}

Due to the extra mass-loss channels (stellar evolution and
evaporation) and the resultant reduction in the inspiral (and hence
tidal stripping) rate, the lifetimes of the clusters shown in
Fig.\,\ref{Fig:RtNXk} may be either longer or shorter than those of
clusters in which stellar evolution is neglected (as in
Fig.\,\ref{Fig:F2}).  This is illustrated in Fig.\,\ref{Fig:6} for
models having $\mu = 0.00012$ and $\beta = 0.05$, 0.025 (the
``standard'' model) and 0.0125.  Dimensionless times are converted to
megayears using an orbital period of 0.30\,Myr, appropriate to a
cluster at an initial distance of 8\,pc from the Galactic center.  The
dotted curves show the evolution of the dimensionless models in which
only tidal stripping is included.  The solid, dashed and dash-3-dotted
curves present the same models with stellar evolution and evaporation
taken into account.  At a distance of 8\,pc from the Galactic center,
the values of $\beta = 0.05$, 0.025, and 0.0125 correspond to
$b_0=0.4$\,pc, 0.2 and $b_0=0.1$\,pc, respectively.

Compact clusters ($b_0\aplt 0.4$\,pc) are relatively unaffected by
tidal stripping.  As a result, the primary effect of stellar mass loss
is simply to decrease the inspiral rate, increasing the cluster
lifetime.  However, in larger clusters ($b_0\apgt 0.4$\,pc) the
expansion caused by stellar mass loss greatly increases the stripping
rate, significantly decreasing the lifetime despite the slower
inspiral.  For very low-mass, or very large, clusters relaxation and
evaporation may dominate.  However for the cases studied here the
effect is almost negligible.  This is illustrated by the first few
million years of evolution of clusters with large $b_0$, such as the
dash-3-dotted curve in Fig.\,\ref{Fig:6}.  The small deviation of the
$b_0 = 0.8$\,pc curve from the $\beta = 0.050$ dotted curve during the
first $\sim5$\,Myr is the result of relaxation.

Fig.\,\ref{Fig:5} illustrates how varying the cluster initial mass
function alters the time evolution of its Galactocentric radius and
mass.  The dotted curve (Fig.\,\ref{Fig:5}a only) shows the constant
point-mass case for $\mu=0.0012$.  The solid curve shows the evolution
of the standard model with $\beta = 0.025$, scaling times to megayears
assuming an initial Galactocentric distance of 8\,pc.  The dashed
curves give the results when stellar mass loss and evaporation are
taken into account, assuming lower mass limits for the initial mass
function of 0.1\,{\msun}, 0.2\,{\msun}, 0.4\,{\msun} and 0.8\,{\msun}.

\begin{figure}[htbp!]
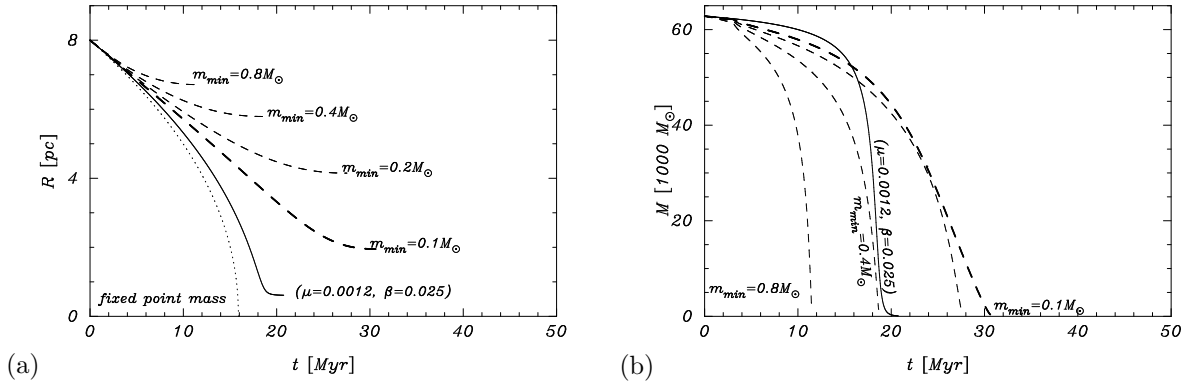

(a)\psfig{figure=f5a.ps,width=7.cm,angle=-90}~~~~~
(b)\psfig{figure=f5b.ps,width=7.cm,angle=-90}
\caption[]{Evolution of (a) the distance to the Galactic center and
(b) mass for model clusters with initial mass $m_0=64,000\,\msun$,
initial radius $b_0=0.2$\,pc, and initial Galactocentric distance
$R_0=8$pc.  The dotted curve in (a) shows the evolution of the model
without stellar evolution or evaporation, assuming that the cluster is
a point mass.  The solid curves are the ``standard'' model, computed
without stellar evolution or evaporation, with $\beta = 0.025$,
appropriate to the choice of $b_0$.  The dashed curves include stellar
evolution and evaporation and are computed using a Salpeter initial
mass function with different lower mass limits {\mmin}, ranging from
0.1\,{\msun} (heavy dashed line) to 0.8\,{\msun} (as indicated).  As
before, we assume $\chi\ln\Lambda = 1$. }
\label{Fig:5}
\end{figure}

Increasing the low-mass cutoff in the mass function increases the
effective stellar evolution mass-loss rate, and reduces the cluster
lifetime.  A similar effect can be achieved by increasing the
power-law slope $x$ of the mass function.  The models are therefore
degenerate in the $x$--$m_{\rm min}$ plane.  For example, a model with
a Salpeter initial mass function ($x = 2.35$) and a low-mass cutoff at
0.2\,{\msun} evolves almost identically to a model with $x = 2.13$ and
$m_{\rm min} = 0.1$ or with $x = 2.90$ and $m_{\rm min} = 0.4$.

\subsection{Comparison with Kim et al.\,(2000)}
Kim (2000)\nocite{2000AAS...197.0409K} used GADGET, the tree code
developed by Springel, Yoshida \& White (2000), to compute the
dynamical friction of dense star clusters near the Galactic center.
In these calculations, the inner part of the Galaxy was represented by
2 million point particles distributed as a truncated softened
power-law similar to our Eq.\,(\ref{Eq:rho}), except that the overall
density was 2.5 times smaller than ours.  The black hole in the
Galactic center was represented as a single particle.  The star
cluster was modeled as a Plummer sphere with $b_0=0.85$\,pc, using
$10^5$ point particles having a total mass of $m_0=10^6$\,\msun.
Initially, the cluster was placed in a circular orbit at a distance of
$R_0=30$\,pc from the Galactic center.  The simulations ignored mass
loss by stellar evolution and evaporation.

\begin{figure}[htbp!]
\psfig{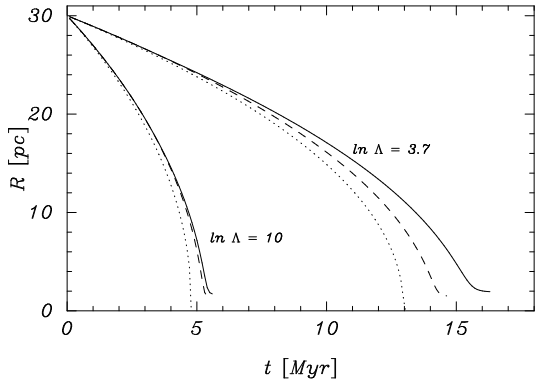}
\caption[]{Time evolution of Galactocentric distance $R$ of a star
cluster having an initial mass $m=10^6$\,\msun.  The rightmost set of
curves represents model 1 of Kim (2000), and have $\ln\Lambda=3.7$;
the leftmost curves have $\ln\Lambda=10$.  Dotted lines are for a
point-mass ($\beta = 0$) cluster with $\mu = 0.00845$.  Dashed lines
are for models with $b_0=0.85$\,pc, but excluding stellar evolution
and evaporation, as assumed by Kim (2000).  Solid curves include
stellar mass loss and evaporation, with a Salpeter initial mass
function between 0.1\,{\msun} and 100\,\msun.  Although Kim (2000)
continued his calculation for only about 9\,Myr, his curve and our
rightmost dashed curve are virtually indistinguishable.  }
\label{Fig:Kimetal}
\end{figure}

The time-dependence of the cluster's Galactocentric distance, as
determined by Kim (2000), is shown in Fig.\,\ref{Fig:Kimetal}.  His
cluster orbits become slightly eccentric during the evolution, but
this seems to have little effect on the dynamical friction timescale.
For clarity we do not show the actual results reported by Kim, but
instead match his initial conditions, which are plotted in
Fig.\,\ref{Fig:Kimetal} as the rightmost dashed curve.  This model is
computed without stellar evolution or evaporation, as in Kim's
simulations.  Our model closely reproduces Kim's results when we adopt
$\ln\Lambda = 3.7$, which is close to the value used by Kim.  For
reference, we also plot the evolution of a constant point-mass model
(dotted curve), and a model in which stellar evolution and evaporation
are taken into account (solid curve).  To guide the eye we also plot
the same series of runs with $\ln \Lambda = 10$.

\section{Discussion}\label{sect:discussion}

In this section we discuss some consequences of our semi-analytical
calculations.  In particular, we consider Gerhard's (2002) conjecture,
discussed in \S1, that IRS\,16 and the associated young stars observed
in the Galactic center may have been deposited there by the inspiral
and disruption of a much more massive system.  We take the two known
Galactic center clusters---the Arches and Quintuplet---as templates.

Table.\,\ref{Tab:Arches} presents the observed parameters for the
Arches and Quintuplet systems.  The final columns give the clusters'
half-mass relaxation time and the time required to reach the Galactic
center, according to Eq.\,(\ref{Eq:trlx}).  (The half mass relaxation
time is computed by substituting {\rhm} for {\rJ} in that equation.)
It is clear that neither cluster will reach the Galactic center within
the next few megayears, and that both were probably born at roughly
their present distance from the Galactic center.  For these
calculations we have again adopted $\chi\log\Lambda = 1$.

\begin{table*}[htbp!]
\caption[]{Observed parameters for the Arches and Quintuplet star
clusters.  Both lie within 35\,pc (in projection) of the Galactic
center.  The first two columns give the cluster name and references,
followed by the distance to the Galactic center, age, mass and half
mass radius.  The last three columns give the two-body relaxation time
at the half-mass radius, the expected time to disruption and the
inspiral time scale.  }
\begin{flushleft}
\begin{tabular}{ll|rrrl|ccc} \hline
Name      &ref.& Rgc&  Age&       M& Rhm & \trlx & \tdiss & \tdf \\
          &    &[pc]&[Myr]& [\msun]&[pc] & 
	\multicolumn{2}{c}{--- Myr ---}  & [Gyr] \\
\hline
Arches    &   a&  30& 2--4&  12--50& 0.2 & 12    &   60   & 0.3--1.9 \\
Quintuplet&   b&  35& 3--5&  10--16& 0.5 & 12    &   60   & 5.5--9.6\\
\hline
\end{tabular} \\
\smallskip
References:
a) Figer et al.\, (1999a);\nocite{1999ApJ...525..750F}
b) Glass, Catchpole \& Whitelock (1987);\nocite{1987MNRAS.227..373G} 
   Nagata et al.\, (1990);\nocite{1990ApJ...351...83N}
   Figer, Mclean \& Morris (1999b).\nocite{1999ApJ...514..202F} 
\smallskip
\end{flushleft}
\label{Tab:Arches} 
\end{table*}

\begin{figure}[htbp!]
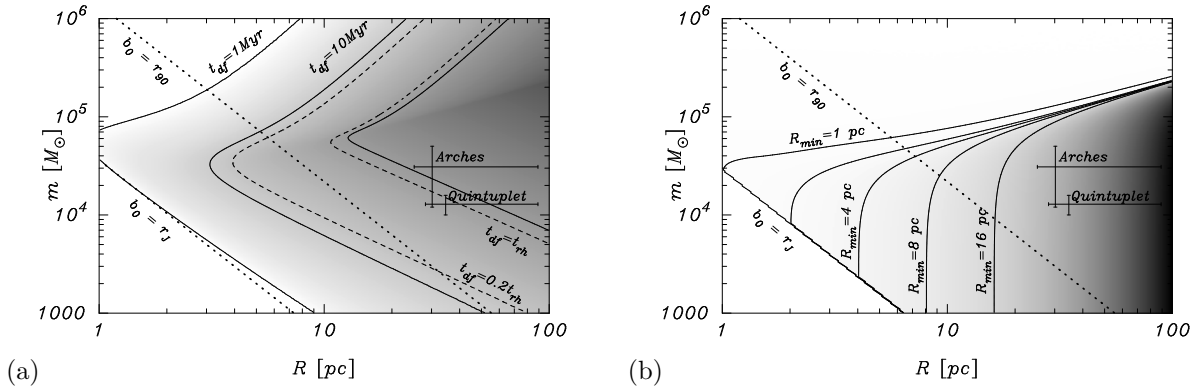

(a)\psfig{figure=f7a.ps,width=7.cm,angle=-90}~~~~~~
(b)\psfig{figure=f7b.ps,width=7.cm,angle=-90}
\caption[]{(a) Inspiral time and (b) final distance to the Galactic
center for clusters with $b_0=0.2$\,pc, as functions of initial
galactocentric distance and cluster mass.  The dotted lines correspond
to $b_0 = 0.9 \rJ$ and $b_0 = \rJ$, as in Fig.\,\ref{Fig:au_cnt}.  The
solid curves indicate initial conditions where the dynamical friction
inspiral time scale $\tdf$ is 1\,Myr (left), 10\,Myr, and 100\,Myr
(right).  The dashed curves in panel (a) correspond to $\tdf = 0.2
t_{\rm rh}$ (left) and $\tdf = t_{\rm rh}$ (right), where $t_{\rm rh}$
is the initial half mass relaxation time of the cluster, obtained by
substituting {\rhm} for {\rJ} in Eq.\,(\ref{Eq:trlx}).  The
approximate locations of the Arches and Quintuplet clusters (see
Tab.\,\ref{Tab:Arches}) are also shown.  }
\label{Fig:7}
\end{figure}

Fig.\,\ref{Fig:7} presents, as a function of initial cluster mass and
galactocentric radius, the time taken for a star cluster with
$b_0=0.2$\,pc to reach the Galactic center (a), and the distance from
the Galactic center at which the cluster dissolves (b).  Contours and
greyscale represent inspiral time in (a) and dissolution distance in
(b).  The cluster is deemed to have dissolved when it comes within
1\,pc of the Galactic center, or when it has lost 99\% of its initial
mass.  The dotted lines have the same meanings is in
Fig.\,\ref{Fig:au_cnt}.

The rightmost dashed line in Fig.\,\ref{Fig:7} marks initial
conditions for which the inspiral time scale equals the initial
relaxation time.  To the right of this curve, the cluster will
experience significant internal dynamical evolution before disrupting.
Our simple description of the cluster's internal structure is
therefore unreliable to the right of this curve, but our expression
for the evaporation rate is still valid.  The left dashed curve
corresponds to an inspiral time scale of 0.2\,\trxh, where {\trxh} is
the initial relaxation time at the half-mass radius.  This is roughly
the core-collapse time for a system with a realistic initial mass
function in which stellar evolution is relatively unimportant (see
Portegies Zwart \& McMillan 2002).\nocite{2002ApJ...576..899P}
Clusters with initial conditions to the left of the left dashed curve
are thus expected to dissolve in the Galactic tidal field before
experiencing core collapse.

During and after core collapse (to the right of the left dashed curve
in Fig.\,\ref{Fig:7}a) the structure of the cluster changes
considerably, and our simple prescription for cluster disruption is
unlikely to hold.  We expect that the structural changes in these
clusters will cause their dense cores to survive for longer, and that
they will sink slightly closer to the Galactic center than indicated
in Fig.\,\ref{Fig:7}(b) (see Gerhard 2002 and Portegies Zwart et al
2003 for further discussion).  The change in disruption radius is not
expected to be great, however, as the residual core masses are small
and their inspiral correspondingly slow at late times.

Fig.\,\ref{Fig:7} clearly indicates that the two known nuclear star
clusters, the Arches and Quintuplet, will not reach the Galactic
center.  Their survival times are determined by internal relaxation
rather than by dynamical friction (see also Tab.\,\ref{Tab:Arches}).
The inspiral time scale for these clusters is of the order of 1 Gyr,
compared to their predicted lifetime of 100\,Myr, based on {\nbody}
simulations (Portegies Zwart et al.,
2001).\nocite{2001ApJ...546L.101P} Fig.\,\ref{Fig:7}(b) indicates that
clusters with $M\aplt 20,000$\,{\msun} barely evolve in
Galactocentric radius on this time scale, but instead dissolve in
situ, due to the combined effects of evaporation and stellar mass
loss.

\begin{figure}[htbp!]
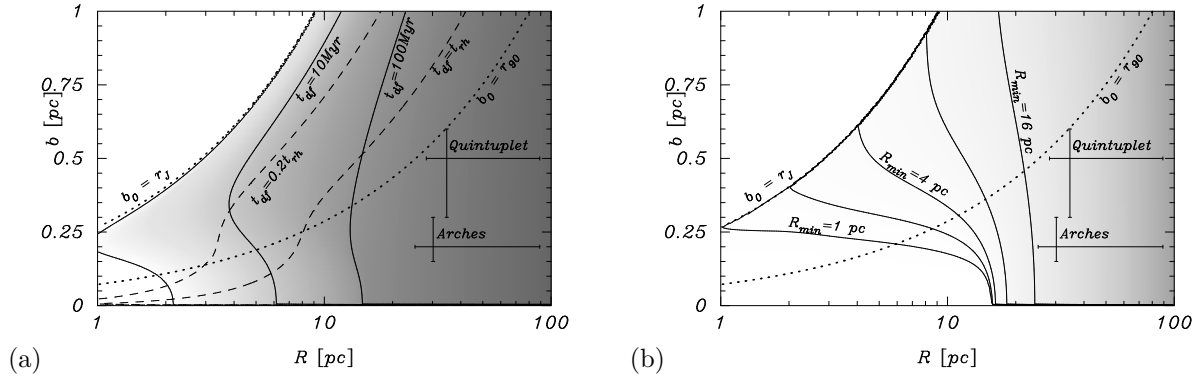

(a)\psfig{figure=f8a.ps,width=7.cm,angle=-90}~~~~~~
(b)\psfig{figure=f8b.ps,width=7.cm,angle=-90}
\caption[]{Contours and greyscale map of (a) inspiral time and (b)
final distance to the Galactic center for clusters with
$m_0=64,000$\,{\msun}, as functions of initial Galactocentric distance
$R_0$ and cluster size $b_0$.  The dashed and dotted curves have the
same meanings as in Fig.\,\ref{Fig:7}, and the two crosses mark the
estimated locations of the Arches and Quintuplet clusters (see
Tab.\,\ref{Tab:Arches}).  }
\label{Fig:8}
\end{figure}

Fig.\,\ref{Fig:8} gives the time taken for a star cluster with
$m_0=64,000$\,{\msun} to reach the Galactic center and the
Galactocentric distance at which the cluster dissolves.  In
Fig.\,\ref{Fig:7} the initial cluster mass $m_0$ was varied at
constant $b_0$.  Now we vary $b_0$ keeping $m_0$ constant, thereby
providing a second slice through the same parameter space as in
Fig.\,\ref{Fig:7}.  As in Fig.\,\ref{Fig:7}, the approximate locations
of the Arches and Quintuplet clusters are indicated.  It is clear that
small shifts in either figure will not alter the basic conclusion that
both clusters will dissolve at large distances from the Galactic
center.

From Figs.\,\ref{Fig:7} and \ref{Fig:8} it is clear that only massive
($\apgt 10^5$\,\msun) star clusters can transport a significant
fraction of their mass to the vicinity of the Galactic center within a
few megayears.  Also, even a million solar mass star cluster will
require several tens of megayears to reach the Galactic center from an
initial distance of $\apgt 30$\,pc.  The most promising candidates to
reach the central parsec of the Galaxy within 10\,Myr, but after
significant mass segregation has occurred, are star clusters with
masses $\aplt10^5$\,{\msun}, born within about 20\,pc of the Galactic
center, with half mass radii of $\sim0.2$--0.4\,pc.  Less massive
clusters, clusters farther from the Galactic center, or smaller
(larger) clusters have greater difficulty reaching the Galactic center
before disruption (core collapse).

We therefore conclude that, if they originated in a massive star
cluster, the stars in IRS\,16 were born in a $\aplt 10^5$\,{\msun}
cluster at a Galactocentric distance of $\aplt 20$\,pc.  The cluster
deposited about $10^3$\,{\msun} of material within $\sim3$\,pc of the
Galactic center.  Since such a cluster would have experienced core
collapse on about the same time scale, the most massive stars had
already segregated to the cluster core.  The deposited (core) material
was therefore rich in massive stars.  These findings are contrary to
the results reported by Kim et al.\,(2002).

More detailed studies are underway to qualify and quantify these
statements (Portegies Zwart, McMillan \& Gerhard 2003).  Preliminary
results indicate that the inspiral times derived here are in good
agreement with {\nbody} calculations using the GRAPE-6 special-purpose
computer (Makino et al 2002), with the same description of dynamical
friction as presented here.  More sophisticated calibration of the
dynamical friction parameters themselves, obtained by modeling the
Galactic background as individual stars, will be the subject of a
future paper.

\acknowledgements

We are grateful to Jun Makino and Piet Hut for helpful discussions,
and thank the Institute for Advanced Study and Tokyo University for
their hospitality and the use of their GRAPE-4 and GRAPE-6 hardware.
This work was supported by NASA through Hubble Fellowship grant
HF-01112.01-98A awarded by the Space Telescope Science Institute and
by NASA ATP grants NAG5-6964 and NAG5-9264, by the Royal Netherlands
Academy of Sciences (KNAW) and the Netherlands Research School for
Astronomy (NOVA).



\section*{Appendix A}\label{App:A}
The argument $X = \vorb/\sqrt{2}\sigma$ in the dynamical friction
relation (Eq.\,\ref{Eq:afric}) of Sec.\,\ref{sect:dynfric} may be
evaluated as follows for inspiral through a sequence of nearly
circular orbits.

Following Binney \& Tremaine (1987, Eq.\,4-30), we write the equation
of dynamical equilibrium (the radial Jeans equation) for stars near
the Galactic center as
\begin{equation}
	\frac{d~}{dR}(\rho\sigma^2) = -\rho\frac{d\phi}{dR}
				    = -\rho\,\frac{\vorb^2}{R}\,,
\label{Eq:dyn_eq}\end{equation}
where we assume an isotropic velocity distribution.  In the power-law
region, $M \propto R^\xgc$ (Eq.\,\ref{Eq:Galaxymass}), we further
assume that $\sigma^2\propto\vorb^2$.  It then follows that
$\sigma^2\rho \sim R^{2\xgc-4}$, so
\begin{equation}
	R\,\frac{d~}{dR}(\sigma^2\rho) = (2\xgc-4)\sigma^2\rho\,.
\end{equation}
Substitution in Eq.\,(\ref{Eq:dyn_eq}) then yields $X =
\sqrt{2-\xgc}$.  We note that, as $\xgc\rightarrow1$, this reduces to
the correct expression for an isothermal sphere (see Binney \&
Tremaine, p. 230).

\end{document}